\documentclass[prd,preprint]{revtex4-2}

\usepackage[margin=1in]{geometry}
\usepackage[utf8]{inputenc}
\usepackage{amsmath}
\usepackage{amsfonts}
\usepackage{graphicx}
\usepackage{booktabs}
\usepackage{setspace}
\usepackage{caption}
\usepackage{natbib}
\usepackage{float}
\usepackage{hyperref}
\usepackage{orcidlink}
\usepackage{microtype}
\usepackage{color}
\usepackage[normalem]{ulem}
\usepackage{tikz}
\usetikzlibrary{positioning,shapes.geometric,arrows.meta,calc}

\begin{document}

\title{A Graph Attention Network Framework for Simultaneous Prediction of Galaxy and Dark Matter Halo Properties in GAMA DR4}

\author{Sandeep Aashish \orcidlink{0000 0003 1316 5966}}
\email{aashish@iitp.ac.in}

\affiliation{Department of Physics, Indian Institute of Technology Patna, Patna, Bihar 801106, India}

\author{Abhay V Hegde}
\email{abhay\_2301ai55@iitp.ac.in}

\affiliation{Department of Computer Science and Engineering, Indian Institute of Technology Patna, Patna, Bihar 801106, India}

\author{Aditya Prakash Gupta \orcidlink{0009-0006-0500-0677}}
\email{aditya\_2301ph03@iitp.ac.in}

\author{Ashmit Verma \orcidlink{0009-0000-9780-5582}}
\email{ashmit\_2301ph07@iitp.ac.in}

\author{Md Saif \orcidlink{0009-0008-75789524}}
\email{saif\_2421ph04@iitp.ac.in}

\author{Nishanth Kumaran}
\email{nishanth\_2301ph12@iitp.ac.in}

\author{Shivang Rayat \orcidlink{0009-0004-8229-0845}}
\email{shivang\_2301ph17@iitp.ac.in}

\affiliation{Department of Physics, Indian Institute of Technology Patna, Patna, Bihar 801106, India}

\date{\today}

\maketitle

\section*{Abstract}

We develop a graph-based deep learning model, GAMANet, to predict galaxy and halo properties using the GAMA DR4 dataset. Multiple catalogues are combined to construct a unified sample of 19,469 galaxies and derive a set of photometric, spectroscopic, morphological, and environmental features. The galaxy sample is represented as a graph, with galaxies as nodes and physically motivated connections based on spatial proximity and group membership as edges. A Graph Attention Network with GATv2 layers is then used to learn information from the galaxy neighbourhood while assigning different weights to neighbouring galaxies. The model simultaneously predicts halo mass, stellar mass, and a specific star formation rate proxy. On the test set, GAMANet achieves $R^2 = 0.958$, $0.985$, and $0.982$ for halo mass, stellar mass, and SSFR proxy, respectively, with an RMSE of $0.237$ dex for halo mass. These results demonstrate that graph-based learning can extract useful information about galaxy and halo properties directly from observational data while incorporating the environmental relationships encoded in the galaxy graph.

\newpage

\section{Introduction}

The concept of dark matter was introduced by Fritz Zwicky \cite{Zwicky_33}, who showed that the total mass of the Coma cluster is much larger than what can be accounted for by its luminous matter. This idea was later supported by observations of galaxy rotation curves,  which showed that visible matter alone cannot account for the observed dynamics \cite{Rubin_78, Bosma_78}. Since then, multiple lines of evidence have confirmed that dark matter constitutes a dominant component of the matter content of the Universe, although its fundamental nature remains unknown \cite{Young_17}. 
In the standard cosmological framework, dark matter plays a central role in shaping the large-scale structure of the Universe. It clusters into dense regions known as halos, which are connected by filamentary structures and separated by low-density voids. Gas accumulates and cools within these halos, leading to the formation of stars and galaxies \cite{Somerville_15}. Consequently, galaxies reside within dark matter halos that define their environment. Even within this well-established picture, predicting important galaxy properties such as halo mass, stellar mass, and star formation rate remains a challenging problem. 

Understanding the relationship between galaxies and their dark matter halos is important for studying the nature of dark matter and the formation of large-scale structure in the Universe. This connection can be explored in two ways. One approach starts from the properties of dark matter halos and attempts to predict the galaxies that form within them. Such studies are useful for generating realistic galaxy catalogs and understanding galaxy clustering. The opposite approach begins with observed galaxy properties and aims to infer the characteristics of their host dark matter halos, such as halo mass and environment. This is particularly important in observational cosmology, where galaxy surveys are used to place constraints on the abundance and distribution of dark matter halos. In this work, we focus on the latter approach by using observed galaxy information to predict fundamental physical properties associated with their dark matter environment.

A wide range of observational methods has been developed to estimate the masses of dark matter halos and galaxy clusters. These include gravitational lensing measurements, galaxy rotation curves, abundance matching techniques, Sunyaev--Zeldovich observations, X-ray studies of hot gas, velocity dispersion analyses, and the dynamics of satellite galaxies \cite{Mandelbaum_15, Huang_20, Huang_22, SofueRubin_01, Sofue_15, Behroozi_10, Grego_01, Landry_13, Saro_13, WojtakMamon_13, Seo_20}. Each method probes halo mass through different physical processes and observational signatures, providing complementary information about the dark matter distribution in galaxies and clusters.
Many existing approaches for estimating halo masses make use of only a limited subset of the available observational information. For example, abundance matching methods primarily rely on the total stellar mass of a galaxy system, while often neglecting additional information related to galaxy environments and clustering properties. In this work, we aim to develop a framework that can simultaneously utilize a broader range of observational features, including phase-space information, stellar masses, galaxy sizes, and environmental properties, in order to infer the mass of the dark matter halo hosting the galaxies. To achieve this, we employ neural network-based methods, which are well suited for modeling complex and highly nonlinear relationships present in galaxy formation and evolution.

Machine learning techniques have increasingly been applied to the problem of halo mass estimation. Convolutional Neural Networks (CNNs) have been used to predict the dynamical masses of galaxy clusters from observational and simulated data \cite{Ho_19, Ntampaka_19, GuptaReichardt_20, Yan_20, KodiRamanah_20, deAndres_21, KodiRamanah_21}. Other machine learning approaches, including multilayer perceptrons and related algorithms, have also been employed to infer galaxy and subhalo masses from various physical properties \cite{Ntampaka_15, Ntampaka_16, HaiderAbbas_19, Armitage_19, Green_19, Shao_22, vonMarttens_21}. In some studies, halo masses have additionally been estimated directly from galaxy images using CNN-based models \cite{deLosRios_21}.
Several studies have further explored predicting halo properties using information from both halos and galaxy groups \cite{Man_19, CalderonBerlind_19, LucieSmith_20}. Overall, these machine learning methods have demonstrated improved performance compared to many traditional halo mass estimation techniques. However, many existing approaches rely heavily on N-body simulations combined with semi-analytical galaxy formation models rather than fully observational datasets. In addition, some of the input features used in these studies are difficult to measure directly from observations, which may limit their applicability to real survey data.

While many existing machine learning approaches make use of global halo properties and individual galaxy features, they often do not explicitly account for the relationships between galaxies, such as their spatial clustering or phase-space distribution. However, these interactions contain important information about the underlying dark matter environment and the large-scale structure of the Universe.
Galaxy evolution is strongly influenced not only by intrinsic galaxy properties but also by the surrounding large-scale environment. Galaxies interact gravitationally within groups, clusters, filaments, and the broader cosmic web, leading to correlations in their dynamical and evolutionary properties. Consequently, relational information between galaxies may contain important signatures of the underlying dark matter distribution that are not captured by treating galaxies as isolated systems. Graph-based methods provide a natural framework for incorporating these complex interactions into machine learning models \cite{Wu_19, Bronstein_21}.

In this work, we address this limitation using a framework based on Graph Neural Networks (GNNs). Unlike traditional deep learning methods that are designed for regular data structures, GNNs operate naturally on graph-based data where objects can have complex and irregular relationships. In our case, galaxies can be represented as nodes, while their physical and environmental connections define the edges of the graph.
This approach differs from Convolutional Neural Networks (CNNs), which are primarily developed for structured grid-like data such as images, and from recurrent neural networks, which are designed for sequential data such as time series or language processing. Galaxy catalogs, on the other hand, are more naturally described as irregular point distributions with nontrivial relational structure, making graph-based methods particularly suitable.

Graph Neural Networks have already demonstrated strong performance in several scientific and computational fields, including chemistry, computer vision, natural language processing, and particle physics \cite{Wu_19, Bronstein_21, Shlomi_20}. In cosmology, GNN-based approaches have recently been explored for applications such as symbolic regression, galaxy redshift prediction, and unsupervised galaxy selection \cite{Cranmer_19, Cranmer_20, BeckSadowski_19, Cranmer_21}. These studies highlight the potential of graph-based learning as a powerful and still relatively unexplored tool for extracting information from complex astrophysical datasets.
Graph Attention Networks are particularly well-suited for this problem because they allow the model to dynamically assign different levels of importance to neighboring galaxies and their environmental connections. This enables the network to identify the most informative relationships contributing to the prediction of halo and galaxy properties. In addition to predictive performance, attention-based architectures also provide a degree of interpretability by allowing the analysis of learned interaction patterns and latent graph representations \cite{Wu_19,Bronstein_21}.

In this work, we develop a graph-based framework for predicting important galaxy and halo properties directly from observational survey data. Using the GAMA DR4 galaxy catalog, we construct a physically motivated graph representation in which galaxies are treated as nodes and the connections between them encode environmental and relational information associated with the cosmic web. The graph incorporates observationally accessible quantities such as spatial proximity, phase-space structure, stellar properties, and local environmental information. Based on this representation, we employ a Graph Attention Network (GAT)-based architecture using GATv2 layers \cite{Brody_22} to simultaneously predict halo mass, stellar mass, and star formation-related quantities in a multi-task learning framework.
Unlike many previous machine learning approaches that rely primarily on N-body simulations and semi-analytical galaxy formation models, our framework is trained directly on observational galaxy survey data. This is important because simulated datasets may contain modeling assumptions and systematic biases that do not fully capture the complexity of real observations. By using observationally derived galaxy properties and environmental measurements, our approach aims to provide a more directly applicable framework for analyzing large galaxy surveys and studying the galaxy--halo connection in realistic astrophysical environments \cite{Ntampaka_15,Armitage_19,Shao_22,LucieSmith_20}.

The paper is organized as follows. In Section~II, we introduce Graph Attention Networks and outline the proposed framework. Section~III describes the observational dataset, catalogue merging, feature engineering, graph construction, model architecture, and training procedure. Section~IV presents the experimental results and discusses the performance of the proposed model. Finally, Section~V summarizes the main conclusions and discusses the limitations and future directions of this work.


\section{Graph Attention Networks (GAT)}

Graph Neural Networks (GNNs) are a class of machine learning models designed to work with graph-structured data, where objects are represented as nodes and their relationships are represented as edges \cite{Scarselli_09, Gilmer_17, Wu_21}. In astronomy, galaxies do not evolve in isolation but are connected through their spatial distribution, group membership, and surrounding environment. Therefore, representing galaxies as nodes and their physical relationships as edges provides a natural way to describe the structure of the cosmic web and the galaxy--halo connection \cite{Cooray_02, Wechsler_18, Sanchez_23}.

A key feature of GNNs is that each galaxy can update its representation by receiving information from neighbouring galaxies. However, many traditional GNN architectures combine information from all neighbouring nodes using the same weighting scheme \cite{Kipf_17,Hamilton_17}. This assumes that every neighbour contributes equally to the final representation. In astrophysical systems this assumption is often unrealistic. For example, a massive central galaxy may contain more information about the host dark matter halo than a small satellite galaxy, and neighbouring galaxies can be affected differently by environmental processes such as mergers, tidal interactions, and quenching \cite{Wechsler_18,Zheng_07,Robotham_11}.

Graph Attention Networks (GATs) were introduced by \cite{Velickovic_18} to address this limitation. Instead of treating all neighbouring nodes equally, GAT assigns a different importance to each neighbour through an attention mechanism. In this way, the model can automatically learn which neighbouring galaxies are most relevant for a particular prediction task. This is achieved through the a series of operations within each graph attentional layer, which we review below.

A single graph attentional layer is first fed a set of $N$ node feature vectors, $\mathbf{h} = \{\vec{h}_1, \vec{h}_2, \ldots, \vec{h}_N\}$, $\vec{h}_i \in \mathbb{R}^{F}$, where $N$ is the number of nodes and $F$ is the number of features contained in each node. This set of input vectors is first mapped from the original feature space of the nodes into a learned representation space better suited for the attention mechanism to operate on. This is achieved via a learned linear transformation, parametrised by a weight matrix $W \in \mathbb{R}^{F' \times F}$, that is shared across all nodes in the layer. A shared attention mechanism then computes attention coefficients by performing self-attention between nodes. The graph structure is injected into the attention mechanism by restricting this computation to first-order neighbours only and hence, for a node $i$, attention coefficients $e_{ij}$ are computed only with respect to nodes $j$ belonging to its neighbourhood $\mathcal{N}_i$:

\begin{equation}
    e_{ij} = a(W\vec{h}_i, W\vec{h}_j), \qquad j \in \mathcal{N}_i
    \label{eq:gat_raw_score}
\end{equation}

Calculating the attention coefficient between a pair of nodes involves concatenating their transformed feature vectors and taking their weighted sum via a single learned weight vector $\vec{a} \in \mathbb{R}^{2F'}$. This weighted sum is then passed through a LeakyReLU nonlinearity, allowing the attention score to capture non-linear relationships between pairs of nodes. The raw attention scores are finally normalised via a softmax over each node's neighbourhood, making the resulting attention coefficients comparable across different nodes:

\begin{equation}
    \alpha_{ij} = \frac{\exp\left(\text{LeakyReLU}\left(\vec{a}^T
    [W\vec{h}_i \,\Vert\, W\vec{h}_j]\right)\right)}
    {\displaystyle\sum_{k \in \mathcal{N}_i} \exp\left(\text{LeakyReLU}\left(\vec{a}^T
    [W\vec{h}_i \,\Vert\, W\vec{h}_k]\right)\right)}
    \label{eq:gat_alpha}
\end{equation}

\noindent where $\Vert$ denotes concatenation.

The new features of a node are finally computed using the normalised attention coefficients, as a weighted sum of the transformed feature vectors of the node's neighbours, followed by the application of a nonlinearity $\sigma$:

\begin{equation}
    \vec{h}_i' = \sigma\left(\sum_{j \in \mathcal{N}_i} \alpha_{ij} W\vec{h}_j\right)
    \label{eq:gat_aggregate}
\end{equation}

To improve performance, GAT employs multi-head attention, where several independent attention mechanisms operate in parallel. The outputs are then combined to produce the final node representation. This allows the model to learn different aspects of the local graph structure and improves training stability. The new features of a node $i$, computed via $K$ independent attention heads, can thus be represented as follows:

\begin{equation}
    \vec{h}_i' = \big\Vert_{k=1}^{K} \, \sigma\left(\sum_{j \in \mathcal{N}_i}
    \alpha_{ij}^{k} W_k \vec{h}_j\right)
    \label{eq:gat_multihead}
\end{equation}

\noindent where $\Vert$ again denotes concatenation, $\alpha_{ij}^{k}$ are the normalised
attention coefficients computed by the $k$-th attention mechanism ($a^k$), and $W_k$ is
that head's corresponding input linear transformation.

\begin{figure}[t]
    \centering
    \includegraphics[width=\textwidth]{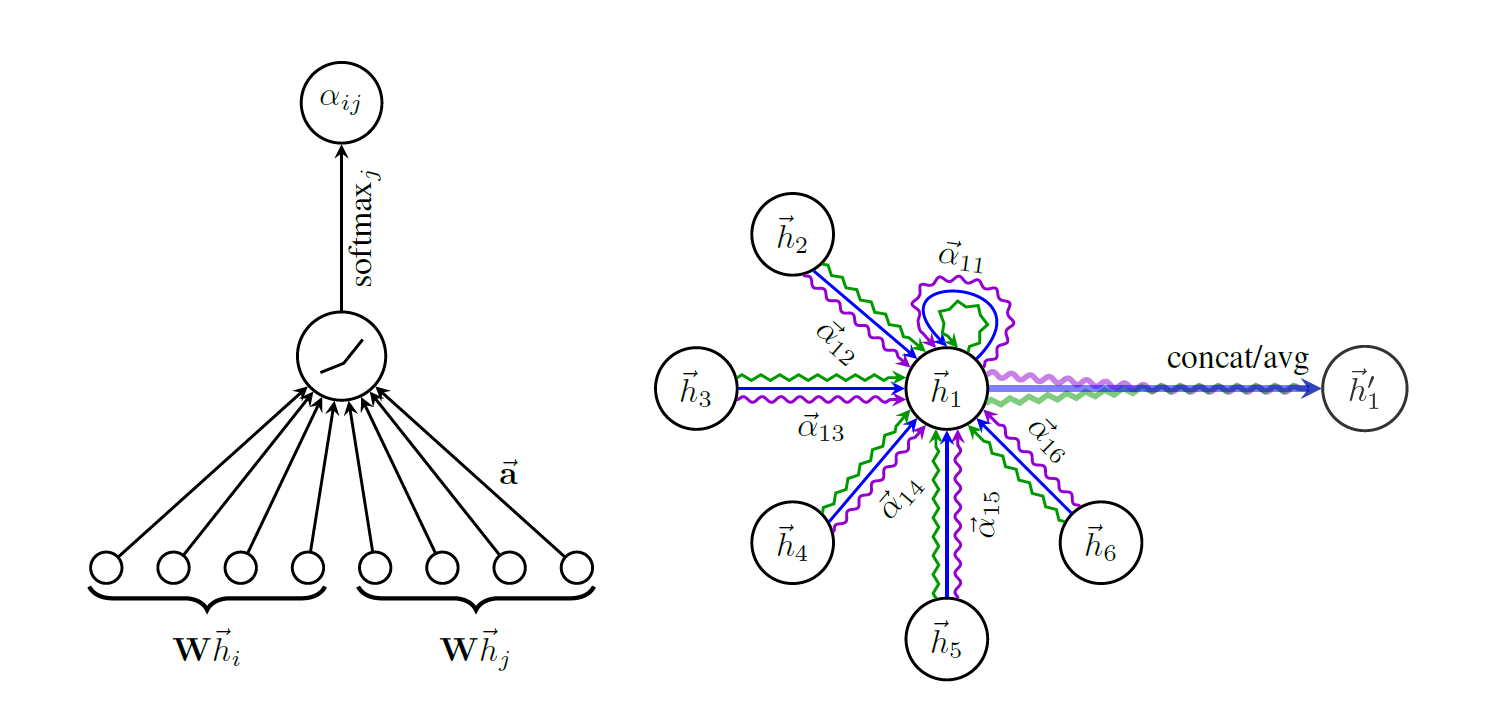}
    \caption{\textbf{Left:} The attention mechanism $a(W\vec{h}_i, W\vec{h}_j)$ employed by
    the GAT layer to yield the attention coefficient $\alpha_{ij}$.
    \textbf{Right:} An illustration of multi-head attention ($K=3$ heads, denoted by the
    green, blue, and purple arrows) as computed by node 1 over its neighbourhood. Each
    colour represents an independent attention mechanism; the resulting per-head features
    are combined via concatenation (in hidden layers)
    to obtain the updated representation $\vec{h}_1'$. Reproduced from Ref. \cite{Velickovic_18}.}
    
    \label{fig:gat_attention_mechanism}
\end{figure}

The attention mechanism makes GAT particularly useful for galaxy studies because the influence of neighbouring galaxies is not uniform. Galaxy properties are known to depend on factors such as local density, group membership, mergers, and large-scale environment \cite{Dressler_80,Peng_10,Wechsler_18}. By assigning larger weights to more informative neighbours and smaller weights to less relevant ones, GAT can better capture these environmental effects than methods based on uniform neighbour averaging.

The objective of this work is to predict three galaxy and halo properties directly from observational data: halo mass ($\log M_{\mathrm{halo}}$), stellar mass ($\log M_{\ast}$), and a specific star formation rate proxy ($\mathrm{SSFR}_{\mathrm{proxy}}$). Rather than treating galaxies as independent objects, we model the observed galaxy distribution as a graph so that both the intrinsic properties of individual galaxies and their relationships with neighbouring galaxies can be incorporated into the learning process.
Our methodology consists of three main steps. First, we merge the GAMA DR4 G02 catalogues into a unified galaxy dataset and extract the relevant galaxy features for training. Second, we construct a graph in which each galaxy is represented as a node, with connections established based on their spatial proximity and group membership. The resulting graph is then processed using a Graph Attention Network (GATv2), which learns from both the intrinsic properties of individual galaxies and their local neighbourhood. Finally, the learned graph representations are used to simultaneously predict halo mass, stellar mass, and the specific star formation rate (SSFR) proxy.


\section{Methodology}

Because no single GAMA catalogue contains all the photometric, spectroscopic, morphological, environmental, and group information required for this work, multiple catalogues were merged into construct a unified dataset. All data used in this work are drawn from the GAMA (Galaxy And Mass Assembly) DR4 G02 field, which covers approximately $55.7 ~deg^2$ of sky. The final dataset was constructed by merging the following 14 FITS catalogues, using either CATAID (the unique GAMA galaxy identifier) or OBJID (the SDSS photometric identifier).

The G02TilingCatv07 catalogue serves as the spectroscopic backbone of the dataset and defines the parent galaxy sample using the quality criteria ($\text{NQ} \ge 3$, $\text{SURVEY\_CLASS} \ge 5$, $\text{DEC} > -6.3$, $0.003 < z < 0.6$). The G02SDSSInputCatv01 catalogue provides the SDSS $ugriz$ photometry and is linked through OBJID, since it does not contain CATAID. The StellarMassesG02SDSSv24 catalogue provides stellar population properties, including MAGPHYS-derived stellar masses, ages, metallicities, and rest-frame colours  \cite{Taylor_11}. Friends-of-Friends (FoF) group membership and halo properties are obtained from the G3CGalv10/G3CFoFGroupv10 catalogues \cite{Robotham_11}, while the G3CGalPairv10/G3CLinkv10 catalogues provide information on physical galaxy pairs and FoF link statistics.

External spectroscopic redshifts are obtained from the ExternalSpecAllv07 catalogue, retaining only measurements with IS\_SBEST = 1 and using the NQ tiebreak for duplicate best measurements. Flow-corrected distances and distance moduli, computed using the flow model of \cite{Tonry_00}, are obtained from the DistancesFramesv14 catalogue. Morphological information from the CFHTLS survey is included through G02CFHTLSInputCatv04, joined using COBJID, with imaging based on a pixel scale of 0.187~arcsec/pixel. Spectral emission-line measurements, including $\text{H}\alpha$, $\text{H}\beta$, [NII] and [OIII] Gaussian fluxes, are taken from GaussFitComplexv05, retaining the spectrum with the highest $\text{H}\alpha$ flux for each galaxy.

Local environmental properties, including projected surface density, nearest-neighbour distances, and survey completeness information, are obtained from EnvironmentMeasuresv06, which covers approximately 33\% of the G02 field. Finally, G02GalacticExtinctionv01/ G02kcorr\_z00v01 provide Galactic extinction corrections and k-corrections, respectively. Together, these catalogues provide a comprehensive observational description of each galaxy, combining photometric, spectroscopic, structural, environmental, and halo-related information for the subsequent graph construction and machine-learning analysis.


\subsection{Catalogue Merging}

During the merging process, several catalogue-specific issues were addressed to ensure a consistent and reliable final dataset.
In the G3CGalv10 catalogue, a value of GROUPID = 0 denotes the absence of a group rather than a valid group identifier. These values were therefore replaced with NaN before merging to prevent isolated galaxies from being assigned incorrect group properties. For the ExternalSpecAllv07 catalogue, the IS\_SBEST = 1 flag is assigned independently within each contributing survey rather than globally. Consequently, 12,505 galaxies contained multiple IS\_SBEST = 1 measurements, which were resolved by selecting the entry with the highest NQ value and, where necessary, the highest PROB value. The GaussFitComplexv05 catalogue contains multiple spectra for many galaxies, yielding 427,829 catalogue entries for approximately 19,000 galaxies. To ensure a unique record for each galaxy, duplicate entries were removed by retaining only the spectrum with the highest H$\alpha$ flux.

In addition, 71 deblended sub-components identified in the G02GalaxyPartMatchesv02 catalogue were removed from the sample before any catalogue merging was performed. Finally, 1,329 guide stars from the G02GuideStarsv01 catalogue and 2,384 flux calibration stars from the G02SpStandardsv01 catalogue were excluded from the analysis. After applying the quality cuts, correcting the \texttt{GROUPID} sentinel values, removing deblended sub-components, and excluding 88 galaxies with unphysical spectral energy distribution (SED) fits ($\log M_{\ast} < 5$ or $\mathrm{SSFR}_{\mathrm{proxy}}$ outside the range $[-3,+3]$ dex), the final sample contains 19,469 galaxies described by 229 merged catalogue columns. This dataset forms the basis for all subsequent feature engineering, graph construction, and machine-learning analyses.


\subsection{Targets}

The model simultaneously predicts three target quantities describing the physical properties of galaxies and their host dark matter haloes. Because the availability of these targets differs across the sample, a masked loss function is employed during training.

The first target, $\mathrm{LOG\_GROUPMASS}$, is defined as $\log_{10}\!\left(M_{\mathrm{halo}}/M_\odot\right)$ using the \texttt{MassA} estimate from the \texttt{G3CFoFGroupv10} catalogue. Halo masses are available only for galaxies belonging to Friends-of-Friends (FoF) groups, corresponding to 7,476 galaxies (38.4\% of the final sample), and are undefined (\texttt{NaN}) for the remaining 12,082 isolated galaxies. In addition, 427 groups with non-positive \texttt{MassA} values were assigned \texttt{NaN}. The second target, $\mathrm{LOGMSTAR}$, is defined as $\log_{10}\!\left(M_{\ast}/M_\odot\right)$ and is obtained directly from the SED-fit parameter \texttt{logmstar} in the \texttt{StellarMassesG02SDSSv24} catalogue. This target is defined for all 19,469 galaxies. The third target, $\mathrm{SSFR\_PROXY}$, is defined as $\log(M_{\mathrm{intSFH}})-\log(M_{\ast})$, where $M_{\mathrm{intSFH}}$ is the total stellar mass integrated over the SED-fitted star formation history. This quantity serves as a mass recycling proxy, with larger values indicating more mass returned to the interstellar medium relative to the retained stellar mass. This target is defined for all 19,469 galaxies.

After defining the prediction targets, a random 70/15/15 split was applied to the galaxy sample to construct the training, validation, and test sets. The split was performed before any scaling or imputation to prevent information leakage from the validation and test sets into the training process. Galaxies with defined $\mathrm{LOG\_GROUPMASS}$ values are approximately uniformly distributed across the three subsets, with 13,628 galaxies (38.1\%) in the training set, 2,920 galaxies (37.2\%) in the validation set, and 2,921 galaxies (38.3\%) in the test set.


\subsection{Feature Engineering}

The merged catalogue was processed to construct the input features used by the Graph Attention Network.
From the 229 merged catalogue columns, 140 engineered features were constructed and organised into seven feature groups. Before feature engineering, columns that could introduce target leakage or contained insufficient observational coverage were excluded. Specifically, the halo-related quantities \texttt{NFOF}, \texttt{VELDISP}, \texttt{RAD50}, \texttt{MEAN\_SEP\_PROJ}, \texttt{MIN\_SEP\_PROJ}, and \texttt{MEAN\_SEP\_RAD} were removed because they are direct inputs to the Friends-of-Friends (FoF) halo mass scaling relation. In addition, \texttt{Z\_EXTSPEC} and \texttt{PROB\_EXTSPEC} were excluded because they are available for only 7.1\% of the sample, for which median imputation would dominate the resulting feature values.

The feature engineering process begins with the construction of photometric features. Three layers of colour are computed for each galaxy: (1) raw observed model magnitudes and colours, (2) extinction-corrected colours subtracting $A_{u/g/r/i/z}$, and (3) rest-frame colours further subtracting k-corrections $\text{KCORR}_{u/g/r/i/z}$. Colour indices $u-r$, $g-r$, $r-i$, $g-z$ are computed at all three correction levels. Physical colour ranges are enforced via clipping (e.g., $g-r$ clipped to $[-0.5, 2.0]$ to remove failed photometric measurements). Additional features include magnitude statistics (mean, std, range across bands), PSF-vs-model concentration ($\text{PSFMAG\_R} - \text{MODELMAG\_R}$), and fibre-vs-Petrosian concentration.

Morphological features are constructed to characterise the physical size and light concentration of galaxies. The physical semi-major axis, $A_{\mathrm{kpc}}$, is computed from the CFHTLS semi-major axis parameter ($A$, measured in pixels) using the cosmology-dependent kpc-per-arcsec conversion for a flat $\Lambda$CDM cosmology ($H_0 = 70~\mathrm{km\,s^{-1}\,Mpc^{-1}}$, $\Omega_m = 0.3$). Galaxy light concentration is quantified using aperture concentration indices derived from the CFHTLS multi-aperture magnitudes, defined as $\mathrm{CONC}_{06\_12} = \texttt{MAG\_APER06\_R} - \texttt{MAG\_APER12\_R}$ and $\mathrm{CONC}_{12\_24} = \texttt{MAG\_APER12\_R} - \texttt{MAG\_APER24\_R}$. The total concentration is then computed as the sum of these two indices.

Spectral features are constructed to characterise the emission-line properties, ionization state, and dust attenuation of galaxies. Log fluxes are computed for H$\alpha$, H$\beta$, N \textsc{II}$\lambda6548$, N \textsc{II}$\lambda6583$, O\textsc{III}$\lambda4959 $, and O\textsc{III}$\lambda5007 $, with a noise floor set at 10\% of the minimum positive flux. BPT diagnostic ratios, $ \log([\mathrm{N \textsc{II}}]/ \mathrm{H} \alpha) $ and $\log([\mathrm{O \textsc{III}}]/ \mathrm{H} \beta) $, are computed to characterise the dominant ionization mechanism and estimate gas-phase metallicity. The Balmer decrement, $\log(\mathrm{H}\alpha/\mathrm{H}\beta)$, is included as a proxy for dust attenuation.

Environmental feature groups include Surface density ($\Sigma$, $\text{Mpc}^{-2}$), the fifth nearest-neighbour distance, and adaptive Gaussian density estimate are log-transformed. An \texttt{ENV\_MASK} binary flag tracks which of the 19,469 galaxies with valid environment measures (33\% coverage). NaN values are preserved, and no zero-imputation is applied to environmental features.

Redshift feature groups include Best redshift, log(1+z), z\textsuperscript{2}, and offsets between redshift frames as a peculiar-velocity proxy. Group/Pair features include central, satellite and brightest-group-galaxy flags, log group rank, log distance from the group centre, log pair count, log link count, and the smallest pair separation. SED features include age/\(\tau\) ratio, log comoving distance.

A two-stage pipeline is applied: (1) median imputation using training-row medians only, to fill NaN values introduced by sparse catalogs; (2) \texttt{StandardScaler} normalisation using training-row mean and standard deviation only. Both the imputer and scaler are fit exclusively on training rows and applied to validation and test rows without refitting. Final feature matrices are \texttt{float32} with zero NaN or Inf values, verified by assertion.


\subsection{Graph Construction}

To capture the spatial and environmental relationships between galaxies, the cleaned catalogue is transformed into a graph representation suitable for Graph Attention Networks (GATs). In this graph, galaxies are represented as nodes, while edges encode physically motivated relationships between them. To represent both local interactions and the large-scale structure of galaxy groups, the graph is constructed in two stages.

The first stage of graph construction identifies physically neighbouring galaxies. A KDTree is built using the three-dimensional Cartesian coordinates computed from the right ascension (RA), declination (Dec), and comoving distance of each galaxy to enable efficient neighbour searches. Galaxy pairs within a comoving radius of $5\,\mathrm{Mpc}$ are connected by an edge. A line-of-sight velocity criterion of $|\Delta v| < 1000~\mathrm{km\,s^{-1}}$, computed as $|\Delta v| = c\,|\Delta z|/(1+\bar{z})$, is then applied as a secondary filter intended to remove chance line-of-sight superpositions of galaxies that are spatially close in projection but physically unassociated. The $1000~\mathrm{km\,s^{-1}}$ value is a conventional choice that brackets the velocity dispersion of massive galaxy clusters ($\sigma \sim 700\text{--}1000~\mathrm{km\,s^{-1}}$), rather than a value calibrated against the G3C group catalogue. In practice, this filter removed no edges from the present dataset; the $5\,\mathrm{Mpc}$ comoving-distance cut is sufficiently restrictive that every pair within it already satisfies the velocity criterion, with a maximum velocity difference of only $340~\mathrm{km\,s^{-1}}$. The velocity criterion is therefore inactive for the G02 sample and is retained for completeness and for generality when the linking radius is enlarged. Each galaxy pair is represented by two directed edges ($i \rightarrow j$ and $j \rightarrow i$) associated with a 12-dimensional edge feature vector. The unsigned edge features are $|\Delta z|$, $\Delta v~(\mathrm{km\,s^{-1}})$, $r_{\mathrm{proj}}~(\mathrm{Mpc})$, $|\Delta(g-r)|$, $|\Delta\log M_{\star}|$, and $|\Delta\log\Sigma|$. The signed edge features are $\Delta(g-r)$, $\Delta\log M_{\star}$, and $\Delta\mathrm{SSFR}_{\mathrm{proxy}}$, with the signs reversed for the opposite edge direction. Two binary features indicate whether the galaxies belong to the same FoF group (\texttt{same\_group}) and whether centrality-rank information is available (\texttt{has\_rank}). The remaining structural feature is the rank product, $R_i \times R_j$, where $R$ denotes the centrality rank.

The second stage of graph construction addresses a critical limitation of the distance-based edges, namely that galaxies belonging to the same Friends-of-Friends (FoF) group can be separated by more than $5\,\mathrm{Mpc}$, particularly in rich, elongated groups, and therefore remain disconnected. To address this, a second set of edges connects all FoF member pairs regardless of their physical separation. These group-aware edges share the same 12-dimensional feature structure as the physical edges, with \texttt{same\_group} always set to 1. An additional thirteenth binary feature, \texttt{is\_physical}, distinguishes KDTree-derived edges (\texttt{is\_physical} = 1) from pure group edges (\texttt{is\_physical} = 0). Where a galaxy pair appears in both edge sets, duplicate edges are merged using \texttt{coalesce()} with mean reduction. The final graph contains 19,469 nodes and 206,128 directed edges, with a mean node degree of approximately 10.6. A total of 2,178 nodes (11\%) remain isolated, having no neighbours within $5\,\mathrm{Mpc}$ and not belonging to any FoF group.

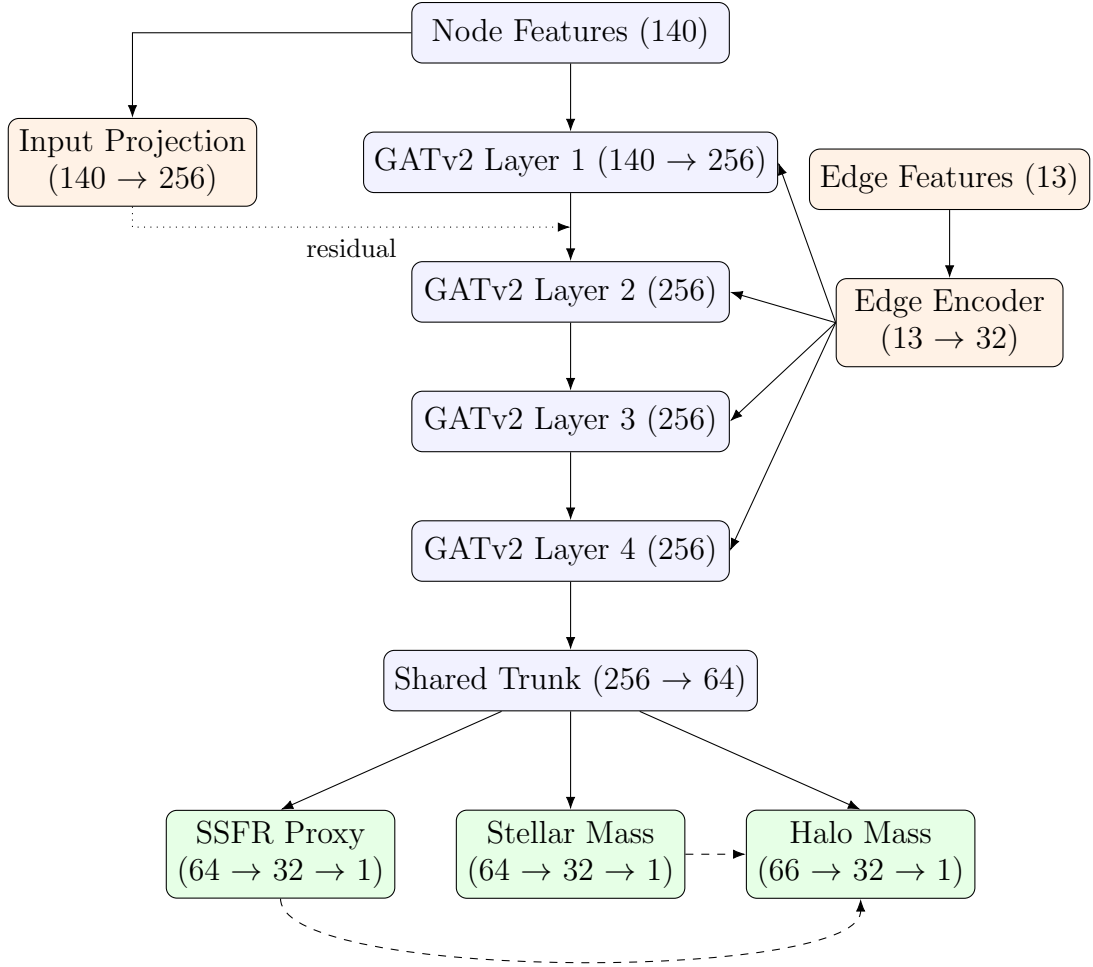
\begin{figure}[t]
\centering
\begin{tikzpicture}[
node distance=0.9cm,
box/.style={draw, rounded corners, minimum width=4.2cm, minimum height=0.8cm, align=center, fill=blue!5},
side/.style={draw, rounded corners, minimum width=3cm, minimum height=0.8cm, align=center, fill=orange!10},
head/.style={draw, rounded corners, minimum width=2.4cm, minimum height=0.8cm, align=center, fill=green!10},
>=Latex
]
\node[box] (input) {Node Features (140)};
\node[box, below=of input] (gat1) {GATv2 Layer 1 (140 $\rightarrow$ 256)};
\node[box, below=of gat1] (gat2) {GATv2 Layer 2 (256)};
\node[box, below=of gat2] (gat3) {GATv2 Layer 3 (256)};
\node[box, below=of gat3] (gat4) {GATv2 Layer 4 (256)};
\node[box, below=of gat4] (shared) {Shared Trunk (256 $\rightarrow$ 64)};
\node[head, below=1.3cm of shared] (stellar) {Stellar Mass\\(64 $\rightarrow$ 32 $\rightarrow$ 1)};
\node[head, left=0.8cm of stellar] (ssfr) {SSFR Proxy\\(64 $\rightarrow$ 32 $\rightarrow$ 1)};
\node[head, right=0.8cm of stellar] (halo) {Halo Mass\\(66 $\rightarrow$ 32 $\rightarrow$ 1)};

\node[side, left=1.4cm of gat1] (proj) {Input Projection\\(140 $\rightarrow$ 256)};
\node[side, right=1.4cm of gat2.south east, anchor=west] (edge) {Edge Encoder\\(13 $\rightarrow$ 32)};
\node[side, above=of edge] (efeat) {Edge Features (13)};

\draw[->] (input) -- (gat1);
\draw[->] (gat1) -- (gat2);
\draw[->] (gat2) -- (gat3);
\draw[->] (gat3) -- (gat4);
\draw[->] (gat4) -- (shared);
\draw[->] (input.west) -| (proj.north);
\draw[->, dotted] (proj.south) |- node[pos=0.75, below, font=\footnotesize]{residual} ($(gat1.south)!0.5!(gat2.north)$);
\draw[->] (efeat) -- (edge);
\foreach \g in {gat1,gat2,gat3,gat4} {\draw[->] (edge.west) -- (\g.east);}

\draw[->] (shared) -- (ssfr.north);
\draw[->] (shared) -- (stellar);
\draw[->] (shared) -- (halo.north);
\draw[->, dashed] (stellar) -- (halo);
\draw[->, dashed] (ssfr.south) .. controls +(0,-1.1) and +(0,-1.1) .. (halo.south);

\end{tikzpicture}
\caption{Architecture of the proposed multi-task graph attention network. The first GATv2 layer takes the raw 140-dimensional node features and the input projection is used only for the first residual connection (dotted), and every subsequent layer adds its input as a residual. The encoded edge features enter the attention computation of every GATv2 layer. Dashed arrows show the stellar-mass and SSFR predictions concatenated with the 64-dimensional shared representation as input to the halo-mass head.}
\label{fig:gnn_architecture}
\end{figure}

The proposed graph neural network architecture consists of three main components: an edge encoder that projects the raw edge attributes into a learned representation, a stack of GATv2 layers that performs message passing over the graph, and a shared multilayer perceptron (MLP) trunk with task-specific output heads for multi-task prediction. The overall architecture is illustrated in Fig.~\ref{fig:gnn_architecture}. The raw 13-dimensional edge feature vectors are projected into a 32-dimensional learned representation using a two-layer multilayer perceptron (MLP) with \texttt{LayerNorm} and \texttt{GELU} activations. This projection normalizes the heterogeneous edge features, such as velocities measured in $\mathrm{km\,s^{-1}}$ and binary indicators, before they are incorporated into the attention computation.

Four \texttt{GATv2Conv} layers \cite{Brody_22} with 8 attention heads each operate on 256-dimensional hidden representations (32 dimensions per head). GATv2 is used over the original GAT because it computes attention as:
\begin{equation}
\alpha_{ij} = a^T \cdot \text{LeakyReLU}(W \cdot [h_i \parallel h_j \parallel e_{ij}])
\end{equation}
which is strictly more expressive than GAT's additive formulation and can distinguish cases where GAT assigns identical attention weights. Each layer includes \texttt{LayerNorm}, \texttt{GELU} activation, 0.2 dropout, and a residual connection. An input projection maps the 140-dimensional node features to the 256-dimensional hidden space for the first residual skip. Learnable per-layer temperature parameters ($T$ initialised at 1.0) allow the softmax attention distribution to sharpen or broaden during training.

The 256-dimensional GATv2 output is passed through a shared MLP trunk ($256 \to 64$) with \texttt{LayerNorm}, \texttt{GELU}, and 0.2 dropout. This shared trunk forces the model to learn a common 64-dimensional representation useful for all three targets, providing implicit multi-task regularisation. Three task-specific output heads ($64 \to 32 \to 1$) produce unbounded predictions for each standardised target, with the halo-mass head additionally receiving the model's stellar-mass and SSFR predictions as inputs. No output activation is applied, as targets are zero-mean unit-variance after scaling. Total trainable parameters: 564,711.

A \texttt{MaskedMultiTaskLoss} module computes a weighted sum of per-target Huber losses. For each target and each training step, the loss is computed only on nodes where: (a) the node belongs to the current split (\texttt{train\_mask}) and (b) the target is defined (\texttt{valid\_mask}). Isolated galaxies contribute to the \texttt{LOGMSTAR} and \texttt{SSFR\_PROXY} losses but contribute zero to the \texttt{LOG\_GROUPMASS} loss. This avoids discarding 62\% of the dataset simply because those galaxies lack halo mass labels. Huber loss deltas are set per-target to match the typical residual scale after standardisation: $\delta = 1.0$ for \texttt{LOG\_GROUPMASS}, $\delta = 0.5$ for \texttt{SSFR\_PROXY}, and $\delta = 0.5$ for \texttt{LOGMSTAR}. Loss weights are equal (1.0 each).

The \texttt{AdamW} optimiser ($\text{lr} = 3\times10^{-4}$, $\text{weight decay} = 10^{-4}$) with \texttt{CosineAnnealingWarmRes-} \texttt{tarts} scheduling ($T_0 = 50$, $T_{\text{mult}} = 2$) is used throughout. Gradient clipping at max norm 1.0 prevents early divergence. Early stopping monitors validation loss with patience of 60--80 epochs depending on the training phase. The best checkpoint is saved on each validation improvement.

Training proceeds in three phases: (1) initial training for 600 epochs with $T_0 = 50$; (2) continuation with longer restart cycles (600 epochs) ($T_0 = 100\text{--}150$) to refine the halo mass head; (3) fine-tuning with a single slow cosine decay (300 epochs) ($\text{lr} = 5\times10^{-5}$ to $10^{-6}$) using the \texttt{resume\_from} checkpoint loading mechanism. The entire graph is held in GPU memory for full-graph transductive training. All nodes participate in message passing every step; \texttt{train/val/test} masks select which nodes contribute to the respective losses.

\section{Results}
Three model versions were evaluated, each introducing structural improvements to the graph. Every version is assessed by the per-target coefficient of determination ($R^2$) computed on the held-out test set, with the root-mean-square error (RMSE, in dex) reported as a secondary metric in Table~\ref{tab:test_performance}. The columns below give $R^2$ for halo mass (GM), the star-formation proxy (SF), and stellar mass (SM):
\begin{table}[htbp]
\centering
\caption{Progressive architecture improvements. All rows use the same 5 Mpc physical graph with group-aware edges; each row adds one design change on top of the previous. Values are test-set $R^2$ for group mass (GM), star-formation proxy (SF), and stellar mass (SM). v4 uses a reduced-capacity backbone (128-dimensional, 3 layers) and is included to show that the hierarchical head requires sufficient backbone capacity to be effective.}
\label{tab:architecture_improvements}
\begin{tabular}{llccc}
\toprule
\textbf{Version} & \textbf{Graph Design} & \textbf{GM $R^2$} & \textbf{SF $R^2$} & \textbf{SM $R^2$} \\
\midrule
v1 & Physical 5 Mpc, 7 unsigned edge features & 0.879 & 0.976 & 0.986 \\
v2 & Physical 5 Mpc, 12 signed edge features + temp. attention & 0.883 & 0.983 & 0.990 \\
v3 & Physical + group-aware edges, 13 features & 0.937 & 0.977 & 0.981 \\
v4 & Hierarchical SHMR head (128-dim, 3 layers) & 0.921 & 0.980 & 0.986 \\
v5 & Hierarchical SHMR head (256-dim, 4 layers) & 0.958 & 0.982 & 0.985 \\
\bottomrule
\end{tabular}
\end{table}

The group-aware edges (v2 $\to$ v3) produced a $+0.054$ improvement in GM $R^2$, the largest single contribution across all design decisions. This confirms that halo mass is fundamentally a group-level property that requires the model to integrate information from all bound members simultaneously, not just spatially proximate neighbours, and supports the inductive bias of a graph-based network: building a denser graph with connections beyond spatial proximity substantially improves performance. The subsequent refinements act primarily on the halo-mass target. Introducing the hierarchical head that feeds the model's own stellar-mass and SSFR predictions into the halo-mass head (v5) raised GM $R^2$ from 0.937 to 0.958, encoding the stellar-to-halo-mass relation directly into the architecture; this gain, however, is only realised at full backbone capacity, as the reduced-capacity v4 underperforms v3 (0.921).

These metrics follow the conventions used in comparable machine-learning studies of the galaxy--halo connection, which typically report $R^2$ together with the scatter (in dex) between predicted and true properties. Existing graph-based halo-mass estimators, however, are trained and validated almost exclusively on cosmological simulations. Villanueva-Domingo \cite{Villanueva_22} train a GNN on more than 2{,}000 CAMELS simulations (using the IllustrisTNG and SIMBA subgrid models), taking galaxy positions, velocities, stellar masses, and radii as inputs, and recover halo masses with $R^2 \approx 0.96$--$0.97$ and roughly $0.2$~dex accuracy. In such simulated settings the true halo masses are known exactly. By contrast, the present model is trained on the real GAMA DR4 G02 sample (19{,}469 observed galaxies over $55.7$~deg$^2$), where halo masses are themselves derived from Friends-of-Friends group finding and carry observational uncertainty. The final (v5) halo-mass performance ($R^2 = 0.958$, RMSE $= 0.237$~dex; Table~\ref{tab:test_performance}) is therefore obtained in a substantially harder regime than simulation-based benchmarks.

The metrics reported here are computed on the held-out test partition defined by the 70/15/15 split: the 15\% of galaxies (2{,}921 objects, of which 38.3\% belong to Friends-of-Friends groups) that are excluded from training and validation and are never seen by the feature imputer or scaler. Metrics are evaluated once, using the best-validation checkpoint, and each target is scored only on the test galaxies for which that target is defined. In particular, \texttt{LOG\_GROUPMASS} is evaluated only over the grouped subset of the test set, since isolated galaxies have no measured halo mass. The final v5 model (group-aware edges, hierarchical SHMR head) achieves the following:

\begin{table}[htbp]
\centering
\caption{Final Test Set Performance}
\label{tab:test_performance}
\begin{tabular}{lccl}
\toprule
\textbf{Target} & \textbf{$R^2$} & \textbf{RMSE (dex)} & \textbf{Notes} \\
\midrule
\texttt{LOG\_GROUPMASS} & 0.958 & 0.237 & Linear baseline R² = 0.49 without the graph \\
\texttt{SSFR\_PROXY}    & 0.982 & 0.019 & Largely set by SED age/\(\tau\) inputs \\
\texttt{LOGMSTAR}      & 0.985 & 0.084 & Auxiliary target for the halo-mass head \\
\bottomrule
\end{tabular}
\end{table}

\begin{table}[H]
\centering
\caption{Model Architecture Summary}
\label{tab:architecture_summary}
\begin{tabular}{lc}
\toprule
Component & Parameters \\
\midrule
EdgeEncoder (2-layer MLP) & 1,632 \\
GATv2Stack (4 layers, 8 heads) & 539,908 \\
SharedTrunk & 16,576 \\
Head: \texttt{LOGMSTAR} & 2,177 \\
Head: \texttt{SSFR\_PROXY} & 2,177 \\
Head: \texttt{LOG\_GROUPMASS} & 2,241 \\
\midrule
Total & 564,711 \\
\bottomrule
\end{tabular}
\end{table}

\subsection{Final Model Result}

The final model (\texttt{GAMANet}) consists of an \texttt{EdgeEncoder}, a \texttt{GATv2Stack}, a shared MLP trunk, and three task-specific hierarchical output heads. The halo-mass head additionally receives the model's own stellar-mass and SSFR predictions as inputs, encoding the stellar-to-halo-mass relation into the architecture. The parameter distribution across these components is summarised in Table~\ref{tab:architecture_summary}, giving a total of 564,711 trainable parameters. The model operates on a graph containing 19,469 galaxy nodes connected by 69,840 edges, where each node is described by 140 input features and each edge by a 13-dimensional feature vector.

The model achieves stable convergence with a minimum validation loss of 0.0380, indicating effective optimization and strong generalization.

\begin{figure}[H]
    \centering
    \includegraphics[width=0.8\textwidth]{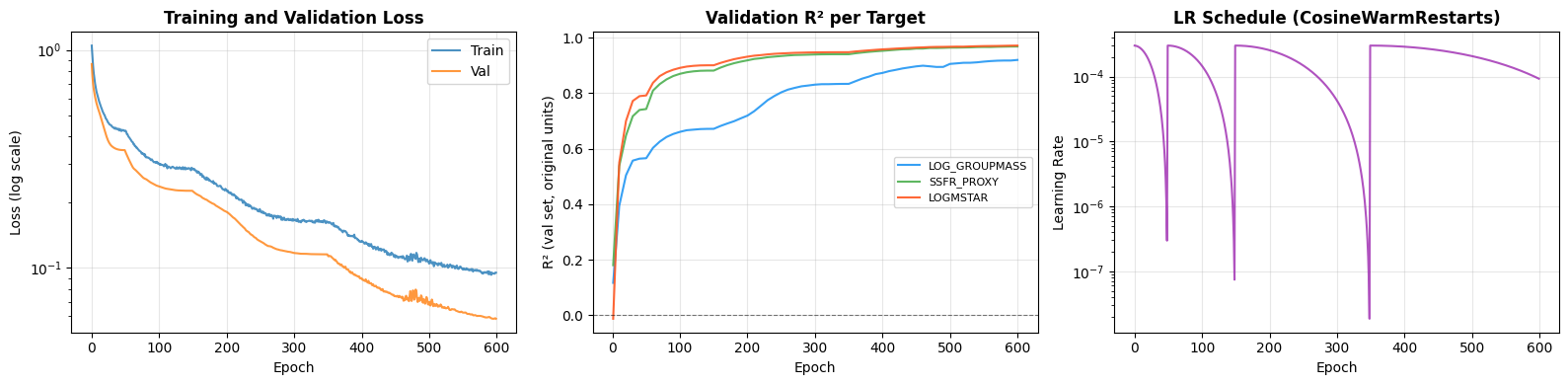}
    \caption{The loss curves show smooth convergence with minimal divergence between training and validation loss, indicating controlled overfitting and stable learning dynamics.}
\end{figure}

The final model demonstrates strong predictive performance across all three targets, achieving $R^2$ values consistently above $0.958$, indicating that a large fraction of the variance is captured across all three targets. The corresponding RMSE values remain low, indicating that the predictions are close to the true values. \texttt{LOGMSTAR} is recovered with $R^2 = 0.985$ (RMSE $= 0.084$~dex). \texttt{LOG\_GROUPMASS} is the hardest target ($R^2 = 0.958$, RMSE $= 0.237$~dex), which is expected: halo mass is a group-level property recovered here without the velocity-dispersion or group-richness features that dominate direct FoF mass estimators, and its labels exist for only the $\sim$38\% of galaxies that belong to groups. 

The progression in Table~\ref{tab:architecture_improvements} clarifies what ``best overall balance'' means. The earlier variants achieve marginally higher stellar-mass and star-formation scores --- v2 reaches SM $R^2 = 0.990$ and SF $R^2 = 0.983$ --- but predict halo mass poorly ($R^2 = 0.879$ and $0.883$ for v1 and v2). The v3 model raises halo-mass $R^2$ to $0.937$, a $+0.058$ gain, at a cost of only about $0.009$ in stellar mass and $0.006$ in star formation. It is therefore preferred not because it is the single best model on every target, but because it removes the large halo-mass deficit of the earlier runs while leaving the other two targets essentially unchanged.

\begin{figure}[H]
    \centering
    \includegraphics[width=0.8\textwidth]{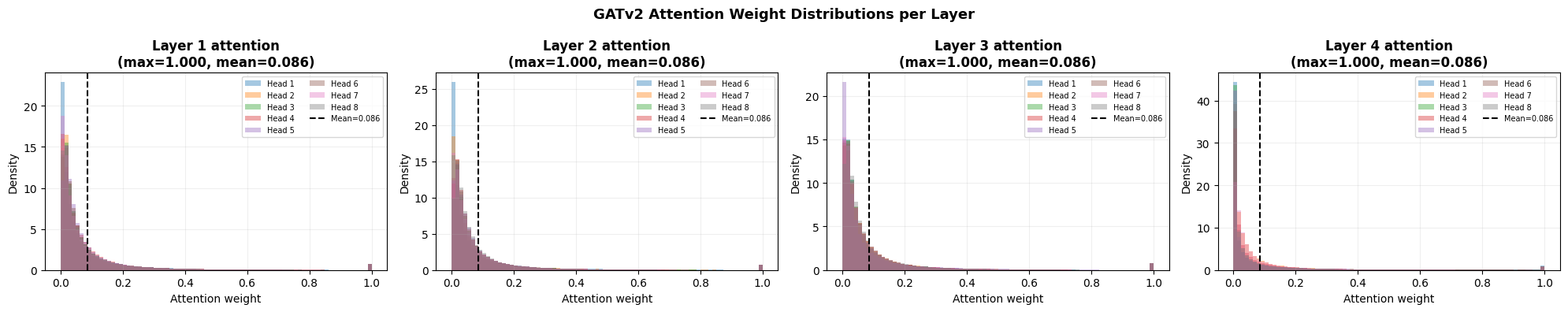}
    \caption{The attention weight distributions across all four GATv2 layers are highly skewed toward low values, with a small fraction of edges receiving high attention weights. The consistent mean attention (~0.086) across layers indicates stable attention behavior. This suggests that the model selectively focuses on a limited number of important neighbors while maintaining sparse interactions, a desirable property in graph attention networks.}
\end{figure}

\begin{figure}[H]
    \centering
    \includegraphics[width=0.8\textwidth]{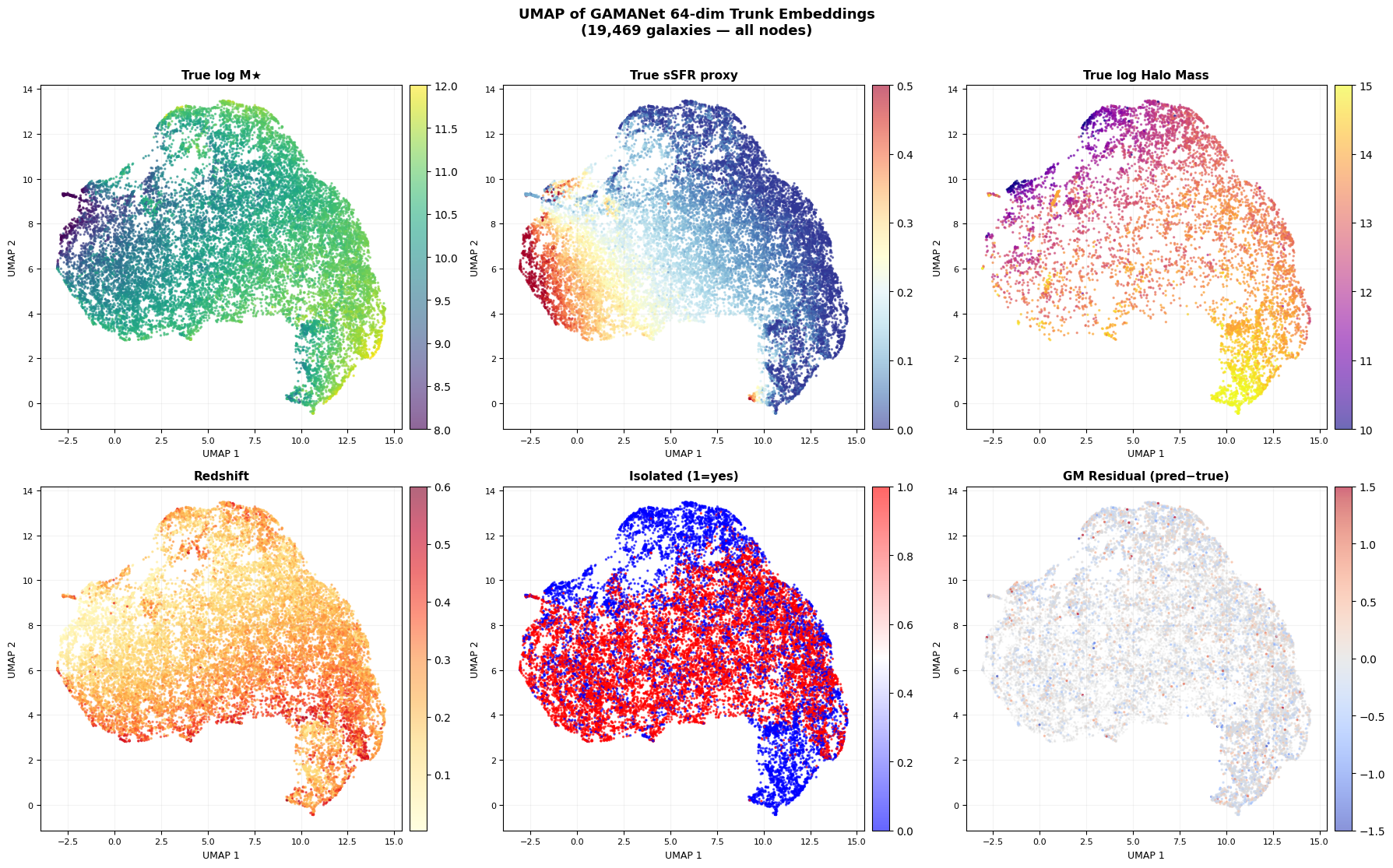}
    \caption{The UMAP projection of the 64-dimensional trunk embeddings reveals a well-structured latent space. Clear gradients are observed for physical properties such as stellar mass, star formation rate proxy, and halo mass, indicating that the model has learned meaningful representations. The smooth transitions across the embedding space suggest that the latent features encode continuous physical relationships between galaxies.}
\end{figure}

\begin{figure}[H]
    \centering
    \includegraphics[width=0.8\textwidth]{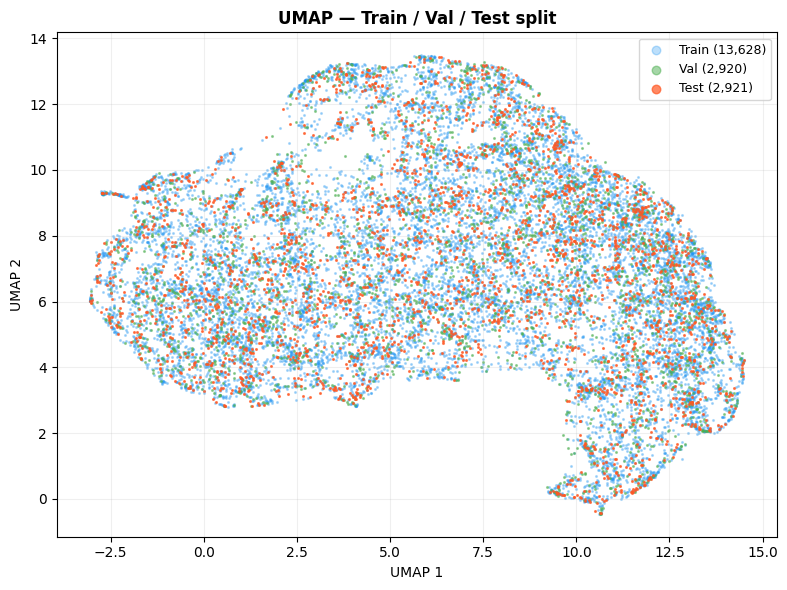}
    \caption{The train, validation, and test samples are well distributed across the embedding space, with no visible clustering or separation between splits. This indicates that the data split is representative and free from sampling bias, supporting reliable generalization performance.}
\end{figure}

\section{Conclusion}

In this work, we developed \texttt{GAMANet}, a Graph Attention Network framework for the simultaneous prediction of dark matter halo mass, stellar mass, and a specific star formation rate proxy using the observational GAMA DR4 G02 dataset. Multiple GAMA catalogues were merged into a single dataset and extracted photometric, spectroscopic, morphological, and environmental features for each galaxy. The galaxies were then represented as nodes in a graph, with edges constructed from their spatial proximity and Friends-of-Friends (FoF) group membership. This graph was processed using a GATv2-based architecture to learn both the properties of individual galaxies and the relationships between neighbouring galaxies.

The final model achieved $R^2$ values of 0.958 for halo mass, 0.985 for stellar mass, and 0.982 for the specific star formation rate proxy. The inclusion of group-aware edges produced the largest single improvement in halo mass prediction, suggesting that halo mass is fundamentally a group-level property. The representation analysis shows that the learned embeddings capture meaningful galaxy properties. The embedding space naturally separates the red sequence from the blue cloud, while the attention analysis shows that physically informative neighbours receive higher attention weights during message passing. In addition, the latent embeddings retain information about stellar velocity dispersion, despite this quantity not being used during training, indicating that the network has implicitly learned galaxy kinematic information from the photometric and environmental features alone.

The model relies on graph attention over one-hop neighbourhoods, and its performance is by how the graph is constructed. The ablation in Table~\ref{tab:architecture_improvements} makes this clear as the largest single gain came not from a change to the network but from adding group-aware edges. This sensitivity to hand-designed connectivity (the $5$~Mpc linking radius, the velocity cut, Friends-of-Friends edges) is a limitation, since the inductive bias is supplied by domain knowledge rather than learned, and a poorly chosen graph would cap performance regardless of model capacity. About $11\%$ of nodes are isolated and receive no message passing at all, so for those galaxies the network reduces to a per-node MLP acting on the input features alone.

Training is full-graph and transductive, the entire graph, together with all intermediate activations, is held in GPU memory, and every node participates in message passing at each step. At about $19{,}500$ nodes this is tractable, but memory scales with the number of nodes and edges, and the transductive formulation means new galaxies cannot be predicted without reconstructing the graph and retraining.  Applying the method at survey scale would require scalable, inductive training for example neighbour sampling (GraphSAGE) or subgraph-based mini-batching (Cluster-GCN, GraphSAINT) which would additionally allow prediction for galaxies absent at training time.

Overall, this work demonstrates that Graph Attention Networks provide an effective framework for predicting important galaxy and halo properties directly from observational survey data. Beyond scalable training, several future extensions follow naturally. Predictive uncertainties would make the halo-mass estimates more useful for downstream cosmology; simulation-based GNN halo-mass estimators already attach uncertainties via likelihood-free inference \cite{Villanueva_22}, and a comparable treatment could be added here. The framework also extends to additional galaxy properties and to the full GAMA footprint and equatorial fields. Testing transferability, training on real data and evaluating on simulations, or the reverse, would directly probe how well a model learned from observations generalises relative to the simulation-trained models that currently dominate this area.

\section*{Acknowledgements}

The authors thank the GAMA collaboration for making the survey data publicly available. We acknowledge the use of opensource libraries in Python, particularly, PyTorch, PyTorch Geometric, Astropy, SciPy, and scikit-learn. UMAP was used for visualization of the learned embeddings.

\section*{Author contribution}
All authors have contributed equally in this manuscript. 


\bibliographystyle{unsrt}
\bibliography{ref}

\end{document}